\documentclass[a4paper]{article}
\usepackage[margin=1.05in]{geometry} % full-width
\usepackage{times,fourier,charter}
\usepackage{graphicx,color}
\usepackage{algorithm,algpseudocode} 
\usepackage{enumitem}
\usepackage{mathrsfs} 
\usepackage[normalem]{ulem} 

\usepackage{amsmath,amsthm,amssymb}
\usepackage{esint}

\usepackage[utf8]{inputenc}
\usepackage[hidelinks]{hyperref}
\hypersetup{unicode,
	pdfauthor={Biondini, Dyachenko, Hoefer and Ossi},
	pdftitle={KPI-Lumps},
	pdfsubject={Modulation theory for lumps and interactions between lumps and a mean field in the Kadomtsev-Petviashvili~equation},
	pdfkeywords={Kadomtsev-Petviashvili equation, Whitham modulation theory, lumps},
	pdfproducer={LaTeX},
	pdfcreator={pdflatex}
}

\makeatletter
\renewcommand\section{\@startsection {section}{1}{\z@}%
  {-2.4ex \@plus -1ex \@minus -.2ex}{1.2ex \@plus.1ex}%
  {\normalfont\bf\sffamily}}
\renewcommand\subsection{\@startsection{subsection}{2}{\z@}%
  {-2.0ex\@plus -0.4ex \@minus -.2ex}{0.8ex \@plus .1ex}%
  {\normalfont\small\bf\sffamily}}
\makeatother

\theoremstyle{plain}

\theoremstyle{definition}

\numberwithin{equation}{section}

\def\@#1{{\mathbf{#1}}}

\def\XXint#1#2#3{{\setbox0=\hbox{$#1{#2#3}{\int}$ }
\vcenter{\hbox{$#2#3$ }}\kern-.575\wd0}}

\graphicspath{{figures/}}

\usepackage{comment}
\usepackage[dvipsnames,svgnames]{xcolor}

\def\gl{\mathrel{\mathpalette\overl@ss>}}

\def\diag{\mathop{\rm diag}\nolimits}

\def\R{\mathbb{R}}

\def\d{\mathrm{d}}

\def\@#1{{\mathbf{#1}}}
\def\_#1{{\mathsf{#1}}}
\let\~=\tilde
\let\==\bar
\let\^=\hat
\let\<=\langle
\let\>=\rangle

\def\u{\bar{u}}
\def\v{\bar{v}}
\def\e{\varepsilon}

\def\be{\begin{equation}}
	\def\ee{\end{equation}}
\def\bse{\begin{subequations}}
	\def\ese{\end{subequations}}

\advance\textheight -4mm
\advance\parskip 2pt
\allowdisplaybreaks

\title{Modulation theory for lumps and interactions \\between lumps and a mean field in the Kadomtsev-Petviashvili~equation}
\author{Gino Biondini$^1$, Sergey Dyachenko$^1$, Mark A. Hoefer$^2$, Nicholas J. Ossi$^{1,\dagger}$}
\date{
    \normalsize{
    \it $^1$ Department of Mathematics, State University of New York, Buffalo, NY 14260\\
    \it $^2$ Department of Applied Mathematics, University of Colorado, Boulder, CO 80309\\[1ex]
    $^\dagger$\rm\href{mailto:nossi@buffalo.edu}{nossi@buffalo.edu}\\[1ex]
    \small\rm
	\today
}}
\begin{document}
\maketitle
%%%%%%%%%%%%%%%%%%%%%%%%%%%%%%%%%%%%%%%%%%%%%%%%%%%%%%%%%%%%%%%%%%%%%
\begin{abstract}
%%%%%%%%%%%%%%%%%%%%%%%%%%%%%%%%%%%%%%%%%%%%%%%%%%%%%%%%%%%%%%%%%%%%%
A (2+1)-dimensional hyperbolic system of four quasi-linear partial differential equations is derived that describes the modulations of lump solutions of the Kadomtsev-Petviashvili I (KPI) equation in the presence of a mean field. 
The system is then shown to satisfy the necessary conditions for integrability of hydrodynamic chains.
Moreover, a suitable reduction of the resulting modulation system is applied to study the interactions between lumps and a rarefaction wave for the mean field.
Precise conditions are derived that describe how the lump parameters change as a result of the interaction, and which in particular determine whether the lump is transmitted through or trapped inside the rarefaction wave. 
The theoretical predictions are compared to direct numerical simulations of the KPI equation, showing excellent agreement.
%%%%%%%%%%%%%%%%%%%%%%%%%%%%%%%%%%%%%%%%%%%%%%%%%%%%%%%%%%%%%%%%%%%%%
\end{abstract}
%%%%%%%%%%%%%%%%%%%%%%%%%%%%%%%%%%%%%%%%%%%%%%%%%%%%%%%%%%%%%%%%%%%%%

%\medskip
%%%%%%%%%%%%%%%%%%%%%%%%%%%%%%%%%%%%%%%%%%%%%%%%%%%%%%%%%%%%%%%%%%%%%
%\tableofcontents
%%%%%%%%%%%%%%%%%%%%%%%%%%%%%%%%%%%%%%%%%%%%%%%%%%%%%%%%%%%%%%%%%%%%%
\medskip	

\section{Introduction}

The Kadomtsev-Petviashvili (KP) equation is a prototypical nonlinear wave equation in 2+1 dimensions. It is a universal model for weakly nonlinear waves in the long wavelength regime and it arises in many different fields of applications ranging from plasmas \cite{FNF2013,KP1970} to water waves \cite{AS1979}, cosmology \cite{GT2006_1,GT2006_2}, ferromagnetics \cite{L2002} and elasticity \cite{BDS2025,E1999}. The KP equation is also the prototypical completely integrable system in 2+1 dimensions \cite{NMPZ1984}, and as such it possesses a deep mathematical structure and a variety of exact solutions.  The KP equation comes in two variants, labeled respectively KPI and KPII. Both variants admit multi-soliton solutions, which are known as line solitons since the solution remains constant along certain directions at space infinity. The initial value problem for the KP equation is also amenable to exact treatment via the inverse scattering transform (IST) \cite{AC1991,AF1983_1,AF1983_2,AS1981}. Over the last 40 years, the IST has been extended to initial conditions that are a small perturbation of an exact soliton solution \cite{BPP2006,BPPP2002,VA2002,W2025}. The IST with nonzero boundary conditions has still not been used effectively to study the behavior of solutions, for example, to determine the long-time dynamics. Moreover, no IST formulation exists at present for more general classes of non-decaying initial conditions.  

For the KPII equation, the recent generalization of Whitham modulation theory \cite{W1965} to systems in 2+1 spatial dimensions \cite{ABW2017} made it possible to study a variety of dynamical problems involving partial solitons and bent solitons, as well as interactions between these objects and a mean flow \cite{RHB2021,RHB2022,RMHB2021}.
In the case of the KPI equation, the situation is different since its line solitons are unstable to transverse perturbations \cite{APS1997,PS1993}. On the other hand, the KPI equation admits stable, fully two-dimensional localized rational solutions called lumps \cite{C2023,CZ2022_1,CZ2022_2,M2015,MZBIM1977,MS1996,SA1979}. More general solutions such as higher-order lumps \cite{AV2002,K1978,P1994,YY2025} and lump chains \cite{LGZZ2021,PS1993} have also been derived and studied. In fact, exact solutions of the KPI equation comprised of line solitons, lumps, and lump chains can be obtained \cite{SZZ2022}, typically using Hirota's bilinear method \cite{H2004}. Note that lumps have recently been observed experimentally for the first time in the realm of nonlinear optics \cite{DPBTC2026}.

A natural question is therefore whether it is possible to formulate a modulation theory for lumps in the same spirit of the one that was developed in \cite{ABW2017} for periodic solutions. In this work we answer this question in the affirmative. 
 Specifically, in section~\ref{s:derivation} we derive a four-component (2+1)-dimensional hyperbolic system that governs modulations of a lump in the presence of a slowly varying mean field. 
 In~section~\ref{s:integrability} we then demonstrate that the resulting system satisfies necessary conditions for integrability and show that it can be diagonalized in certain (1+1)-dimensional reductions. 
Moreover, in section~\ref{s:lumpmeaninteractions} we show how the modulation system can be used to effectively study quantitatively a concrete dynamical scenario, namely an interaction between a lump and a rarefaction wave. We use the modulation equations to predict the slow evolution of the amplitude and velocity of the lump during the interaction, as well as the values of the parameters of a lump that fully transmits through the rarefaction wave.
Finally, section~\ref{s:conclusions} ends this work with a discussion and some concluding remarks.
The appendix contains a discussion of the numerical methods used for the computer simulations.

\section{Derivation of the lump modulation system}
\label{s:derivation}

We consider the KPI equation in evolutionary form
\begin{equation}
\label{e:KP}
    u_{t}+uu_{x}+u_{xxx}-v_{y}=0,\qquad v_{x}=u_{y}.
\end{equation}
The exact single-lump solution of \eqref{e:KP} is given by
\bse
\label{e:lump}
\begin{align}
    u(x,y,t)=\u+U(\eta,\theta),\qquad &U(\eta,\theta)=24\frac{-(\eta+q\theta)^2+a^2\theta^2+3/a^2}{\big[(\eta+q\theta)^2+a^2\theta^2+3/a^2\big]^2},\\
    v(x,y,t)=\v+V(\eta,\theta),\qquad &V(\eta,\theta)=\int_{-\infty}^{\eta}U_{\theta}(\eta',\theta)\,\d\eta',\\
    \eta=x-(\u+a^2+q^2)t + \eta_0,\qquad&\theta=y+2qt+\theta_0. \label{e:lumpphases}
\end{align}
\ese
The vector $(\u,\v)$ represents the background on which the lump is placed, and is arbitrary due to the Galilean invariance of the KP equation. The amplitude $A$ and velocity $\@c$ of the lump are
\begin{equation}
    A=U(0,0) = 8a^2,\qquad\@c=\begin{bmatrix}
        \u+a^2+q^2\\
        -2q
    \end{bmatrix}.
\end{equation}
Summarizing, the solution (2.1) is completely determined by four independent parameters $\bar u$, $\bar v$, $a$, $q$ and its origin $(\eta_0,\theta_0)$. 
Next, we are interested in considering solutions in which all of these parameters are slowly varying,
and we aim to derive a system that governs their modulations, which will allow us to describe the dynamics of a lump subject to a slowly varying mean field. 
To accomplish this, we begin by deriving useful integral identities similar to one derived in \cite{ACEHL2023}  for the KdV equation. Subsequently, we invoke a multiple scale ansatz consisting of a slowly varying lump and mean field, and then follow the lump as a fluid parcel is followed in Lagrangian coordinates.

We begin by introducing the splitting $u=\u+U$, $v=\v+V$, where at this stage all quantities are assumed to depend explicitly on space and time. At a later stage, we will make the assumption that the mean field $\u,\v$ 
 and the lump parameters $a$ and $q$ are slowly varying. Upon substitution, \eqref{e:KP} becomes
\bse
\begin{align}
\label{e:split1}
    U_{t}+UU_{x}+U_{xxx}-V_{y}=-(\u U)_{x}-F[\u,\v],\qquad &F[\u,\v]=\u_{t}+\u\u_{x}+\u_{xxx}-\v_{y},\\
\label{e:split2}
    V_{x}-U_{y}=G[\u,\v],\qquad &G[\u,\v]=\u_{y}-\v_{x}.
\end{align}
\ese
Assuming that $U$ and $V$ are localized, multiplying \eqref{e:split1} by $U$ and integrating over all space gives
\begin{equation}
\label{e:cons1}
    \frac{\d}{\d t}\iint_{\R^{2}}U^{2}\,\d x\,\d y=-\iint_{\R^{2}}U^{2}\u_{x}\,\d x\,\d y-2\iint_{\R^{2}}UF[\u,\v]\,\d x\,\d y.
\end{equation}
If $\u=\v=0$, this simply corresponds to conservation of horizontal momentum. We now aim to derive a similar identity for the transverse momentum. To this end, from \eqref{e:split2} we get
\begin{equation}
    \iint_{\R^{2}}UV_{t}\,\d x\,\d y=\iint_{\R^{2}}U\partial_{x}^{-1}U_{yt}\,\d x\,\d y+\iint_{\R^{2}}U\partial_{x}^{-1}G[\u,\v]_{t}\,\d x\,\d y,
\end{equation}
with a suitable definition of the antiderivative operator. 
Integrating by parts in $y$ in the first term on the right-hand side, substituting $U_{y}=V_{x}-G[\u,\v]$, and then integrating by parts in $x$ in both terms yields
\begin{equation}
\label{e:UV1}
    \iint_{\R^{2}}UV_{t}\,\d x\,\d y=\iint_{\R^{2}}VU_{t}\,\d x\,\d y+\iint_{\R^{2}}\big(G[\u,\v]\partial_{x}^{-1}U_{t}-G[\u,\v]_{t}\partial_{x}^{-1}U\big)\,\d x\,\d y.
\end{equation}
Next, multiplying \eqref{e:split1} by $V$ and integrating we also find
\begin{equation}
\label{e:UV2}
    \iint_{\R^{2}}VU_{t}\,\d x\,\d y+\iint_{\R^2}(VUU_{x}+VU_{xxx})\,\d x\,\d y=-\frac{1}{2}\iint_{\R^2}U^2\u_{y}\,\d x\,\d y-\iint_{\R^2}VF[\u,\v]\,\d x\,\d y.
\end{equation}
Putting \eqref{e:UV1} and \eqref{e:UV2} together and making use of
\begin{equation}
    \iint_{\R^{2}}VUU_{x}\,\d x\,\d y=-\frac{1}{2}\iint_{\R^2}U^{2}G[\u,\v]\,\d x\,\d y,\qquad\iint_{\R^{2}}VU_{xxx}\,\d x\,\d y=-\iint_{\R^{2}}U_{xx}G[\u,\v]\,\d x\,\d y,
\end{equation}
we arrive at the identity
\begin{align}
    \frac{\d}{\d t}\iint_{\R^{2}}UV\,\d x\,\d y=&-\iint_{\R^{2}}U^{2}\u_{y}\,\d x\,\d y-2\iint_{\R^{2}}VF[\u,\v]\,\d x\,\d y\nonumber\\&-\iint_{\R^{2}}G[\u,\v]\big(\partial_{x}^{-1}V_{y}-U\u-\partial_{x}^{-1}F[\u,\v]\big)\,\d x\,\d y-\iint_{\R^{2}}G[\u,\v]_{t}\partial_{x}^{-1}U\,\d x\,\d y.
\label{e:cons2}
\end{align}
Note that for slowly varying $F$, $\partial_{x}^{-1}F$ is not necessarily well-defined, but this term will be of higher order in the multi-scale analysis we now initiate.

We now take $U$ and $V$ to be the lump solution as in \eqref{e:lump} and allow the lump parameters to vary slowly with space and time.
That is, introducing a small parameter $0<\e\ll1$, we define the slow spatial and temporal scales $X=\e x$, $Y=\e y$, $T=\e t$ and
\bse
\begin{gather}
    \u=\u(X,Y,T),\qquad \v=\v(X,Y,T),\qquad a=a(X,Y,T),\qquad q=q(X,Y,T),\\
    \eta=x-\frac{1}{\epsilon}\int_{0}^{\epsilon t}\big[\u(X,Y,T')+a(X,Y,T')^2+q(X,Y,T')^2\big]\,\d T',\qquad\theta=y+\frac{2}{\epsilon}\int_{0}^{\epsilon t}q(X,Y,T')\,\d T'.
\end{gather}
\ese
Next, we introduce the multiple scale ansatz
\begin{equation}
    \label{e:ansatz}
    u(x,y,t)=\u(X,Y,T)+U(\eta,\theta,X,Y,T),\qquad v(x,y,t)=\v(X,Y,T)+V(\eta,\theta,X,Y,T),
\end{equation}
which represents a modulated lump propagating on a slowly varying mean. Accordingly, we have the operator expansions
\begin{equation}
    \partial_{x}=\partial_{\eta}+\e\partial_{X},\qquad\partial_{x}^{-1}=\partial_{\eta}^{-1}(1-\e\partial_{\eta}^{-1}\partial_{X}+\cdots),\qquad\partial_{y}=\partial_{\theta}+\e\partial_{Y},\qquad\partial_{t}=(\u+a^2+q^2)\partial_{\eta}-2q\partial_{\theta}+\e\partial_{T}.
\end{equation}
Furthermore, we enter the reference frame of the lump, following it like a Lagrangian particle in Eulerian coordinates $(X,Y,T)$. Thus, the total time derivative should be interpreted as the convective derivative
\begin{equation}
    \frac{\d}{\d t}=\e(\partial_{T}+\@c\cdot\nabla),\qquad\nabla=\begin{bmatrix}
        \partial_{X}\\\partial_{Y}
    \end{bmatrix}.
\end{equation}
Inserting this ansatz and keeping terms of $\mathcal{O}(\epsilon)$, we find that in this multi-scale setting the generic identities \eqref{e:cons1} and \eqref{e:cons2} provide the following relations:
\bse
\label{e:mod_dkp}
\begin{align}
    (\partial_{T}+\@c\cdot\nabla)\iint_{\R^2}U^2\,\d\eta\,\d\theta+\u_{X}\iint_{\R^{2}}U^2\,\d\eta\,\d\theta&=-2(\u_{T}+\u\u_{X}-\v_{Y})\iint_{\R^2}U\,\d\eta\,\d\theta,\\
    (\partial_{T}+\@c\cdot\nabla)\iint_{\R^2}UV\,\d\eta\,\d\theta+\u_{Y}\iint_{\R^{2}}U^2\,\d\eta\,\d\theta&=-2(\u_{T}+\u\u_{X}-\v_{Y})\iint_{\R^2}V\,\d\eta\,\d\theta+\u(\u_{Y}-\v_{X})\iint_{\R^{2}}U\,\d\eta\,\d\theta.
\end{align}
\ese
Note that due to its slow algebraic decay, the integral of the lump solution $U$ over all space is only conditionally convergent. To remove this ambiguity, the mean field must satisfy the dispersionless KP (dKP) equation:
\begin{equation}
    \label{e:dKP}
    \u_{T}+\u\u_{X}-\v_{Y}=0,\qquad\v_{X}=\u_{Y}.
\end{equation}
The other relevant integrals are well-defined and are found to be
\begin{equation}
    \iint_{\R^{2}}U^{2}\,\d\eta\,\d\theta=96\pi a,\qquad\iint_{\R^{2}}UV\,\d\eta\,\d\theta=96\pi aq.
\end{equation}
Putting these into \eqref{e:mod_dkp} gives the modulation equations for the lump parameters:
\bse
\label{e:mod_conv}
\begin{align}
    (\partial_{T}+\@c\cdot\nabla)a+a\u_{X}&=0,\\
    (\partial_{T}+\@c\cdot\nabla)(aq)+a\u_{Y}&=0.
\end{align}
\ese
In matrix form, the full modulation system including \eqref{e:dKP} and \eqref{e:mod_conv} can be written as
\bse
\label{e:mod_full}
\begin{align}
    \label{e:mod_compact}
    \mathsf{E}_3 \@v_{T}+\mathsf{A}\@v_{X}+\mathsf{B}\@v_{Y}=0,\qquad
    \@v=\big[\,a\,,\, q\,,\, \u\,,\, \v\,\big]^\top,  \\
\noalign{\noindent where the superscript $\top$ denotes matrix transpose, 
$\mathsf{E}_3 = \diag(1,1,1,0)$, and}
    \mathsf{A}=\begin{bmatrix}\u+a^2+q^2&0&a&0\\
    0&\u+a^2+q^2&-q&0\\
    0&0&\u&0\\
    0&0&0&1\end{bmatrix},\qquad&\mathsf{B}=\begin{bmatrix}
        -2q&0&0&0\\
        0&-2q&1&0\\
        0&0&0&-1\\
        0&0&-1&0
    \end{bmatrix}.
\end{align}
\ese
Note how the system~\eqref{e:mod_full} is partially decoupled, since~\eqref{e:dKP} do not depend on $a$ and $q$.
Conversely, \eqref{e:mod_conv} show that the dynamics of the lump \emph{is} influenced by the modulation of the mean field.
The system~\eqref{e:mod_full} is one of the main results of this work.  
In the upcoming sections we study some of the properties of the system and we also show how the system can be used to quantitatively describe the interactions between lumps and a mean field.

It is worth commenting on how the modulation equations~\eqref{e:mod_conv} should be interpreted, since at first sight their PDE form might seem incongruent with a lump being a localized, finite-energy object.  
Indeed, a single lump is described by two quantities $(a,q)$ evolving as a function of $T$ along its trajectory, and one might have therefore expected a system of ODEs for its parameters rather than the PDEs~\eqref{e:mod_conv}.  
The resolution of this apparent paradox is that~\eqref{e:mod_conv} should not be seen as the equations of motion of a single lump, but rather as the equations of motion of a slowly varying \emph{lump field} defined throughout the spatial domain, of which the trajectory of any individual lump is but a single characteristic.  
The dynamics of a single lump are recovered by restricting~\eqref{e:mod_conv} to a characteristic curve
$(X(T),Y(T))$, as will be demonstrated in sections~\ref{s:integrability} and~\ref{s:lumpmeaninteractions}.  
Off this curve, the lump field is to be interpreted as describing a dilute distribution of lumps with the same local parameters.  This is a useful fiction for the purpose of modulation theory, but one that becomes a literal description of the solution if a lump is actually placed at the corresponding point in space.
This interpretation is perfectly analogous to the soliton limit of the Whitham modulation equations for the KdV equation, which is obtained as the $k \to 0$ degeneration of the genus-one KdV-Whitham system and which describes the slow evolution of the parameters of a soliton placed on a slowly varying mean flow~\cite{ACEHL2023,ElHoefer2016}.  
As noted in~\cite{ACEHL2023,MaidenEtAl}, 
the resulting hyperbolic system has the dual interpretation of governing either a single soliton along its trajectory or a dilute soliton field distributed throughout the spatial domain, the two interpretations being consistent because the soliton parameters propagate along the characteristics of the modulation system.  
The system~\eqref{e:mod_full} plays exactly the same role for KPI lumps, with the additional richness that both the lump and its mean-field background are now genuinely two-dimensional.  
At the same time, we should also point out one important difference between the KdV case and the present one: in the former, the soliton-limit modulation system arises as a degenerate reduction of an underlying Whitham system for the periodic cnoidal wavetrain~\cite{ElHoefer2016}, whereas in the present setting equations~\eqref{e:mod_conv} are derived directly from integral identities for the localized lump itself, without any underlying periodic structure.%

\section{Integrability test, integration by characteristics, and two-dimensional reductions}
\label{s:integrability}

Since the KP equation is completely integrable, one expects that
exact asymptotic reductions of it will inherit the integrability properties.  
This was already shown to be true for various reductions of the Whitham modulation equations for the periodic traveling wave solutions of the KP equation in \cite{BHM2020,BBHM2024}.
In this section we show that a similar result applies for the modulation system~\eqref{e:mod_full}.

\subsection{Haantjes tensor test for integrability}

For a (2+1)-dimensional quasilinear system of the form \eqref{e:mod_full}, it was shown in \cite{FK2006}  that a necessary condition for integrability is that the matrix
\begin{equation}
    \mathsf{M}=(k\mathsf{E}_{3}+\mathsf{A})^{-1}(l\mathsf{E}_{3}+\mathsf{B}),
\end{equation}
passes the Haantjes tensor test \cite{H1955} for arbitrary $k$ and $l$. Denoting the entries of $\@v$ in eq.~\eqref{e:mod_compact} as $v^{i}$ and the entries of $\mathsf{M}$ as $M_{j}^{i}$, one first defines the Nijenhuis tensor as
\begin{equation}
    N_{jk}^{i}=M_{j}^{p}\partial_{v^{p}}M_{k}^{i}-M_{k}^{p}\partial_{v^{p}}M_{j}^{i}-M_{p}^{i}\left(\partial_{v^{j}}M_{k}^{p}-\partial_{v^{k}}M_{j}^{p}\right).
\end{equation}
Then, the matrix $\mathsf{M}$ is said to pass the Haantjes tensor test if its Haantjes tensor, defined by
\begin{equation}
    H_{jk}^{i}=N_{pr}^{i}M_{j}^{p}M_{k}^{r}-N_{jr}^{p}M_{p}^{i}M_{k}^{r}-N_{rk}^{p}M_{p}^{i}M_{j}^{r}+N_{jk}^{p}M_{r}^{i}M_{p}^{r},
\end{equation}
is identically zero. Straightforward computations with a computer algebra system demonstrate that this is indeed the case for the system \eqref{e:mod_full}. This suggests that any two-dimensional reduction of \eqref{e:mod_full} can be diagonalized, as will be demonstrated for the $XT$, $YT$, and $XY$ dependent cases in following subsections.
It also suggests that the full system~\eqref{e:mod_full} is also integrable,
as will be also demonstrated in the next subsection.

\subsection{Integration of the lump modulation equations by the method of characteristics}

Recall that the evolution of the mean field is decoupled from the presence of the lump, 
and is governed by the dKP equation~\eqref{e:dKP}, which is completely integrable.
Here we therefore 
fix $\u$ and $\v$ to be a generic solution of the dKP equation~\eqref{e:dKP}. We then write the remaining two modulation equations in the form
\unskip%
\bse
\label{e:mod_fixed_mean}
\begin{align}
    a_{T}+(\u+a^2+q^2)a_{X}-2qa_{Y}=-a\u_{X},\\
    (aq)_{T}+(\u+a^2+q^2)(aq)_{X}-2q(aq)_{Y}=-a\u_{Y}.
\end{align}
\ese
With the mean specified, the method of characteristics can be applied to write \eqref{e:mod_fixed_mean} as a system of four coupled ordinary differential equations for the lump parameters $a(T)=a(X(T),Y(T),T)$, $q(T)=q(X(T),Y(T),T)$ and the characteristic trajectories $X(T)$, $Y(T)$:
\begin{equation}
\label{e:characteristics}
    \frac{\d a}{\d T}=-a\u_{X},\qquad\frac{\d (aq)}{\d T}=-a\u_{Y},\qquad \frac{\d X}{\d T}=\u+a^2+q^2,\qquad\frac{\d Y}{\d T}=-2q.
\end{equation}
It is worth noting that from \eqref{e:characteristics} it is clear that if the mean field is independent of $X$, then $a$ (the horizontal momentum of the lump) is a constant of motion, and if the mean field is independent of $Y$, then $aq$ (the transverse momentum of the lump) is a constant of motion. Furthermore, for any stationary mean $\u=\u(X,Y)$, \eqref{e:characteristics} admits a conserved Hamiltonian 
\begin{equation}
\label{e:Hamiltonian}
    \mathcal{H}(X,Y,a,aq)=a\u(X,Y)+\frac{1}{3}a^{3}-\frac{(aq)^2}{a},
\end{equation}
from which the system is obtained through
\begin{equation}
    \frac{\d a}{\d T}=-\frac{\partial\mathcal{H}}{\partial X},\qquad\frac{\d (aq)}{\d T}=-\frac{\partial\mathcal{H}}{\partial Y},\qquad\frac{\d X}{\d T}=\frac{\partial\mathcal{H}}{\partial a},\qquad\frac{\d Y}{\d T}=\frac{\partial\mathcal{H}}{\partial (aq)}.
\end{equation}
The above equations will be used in a specific situation in section~\ref{s:lumpmeaninteractions}  to describe the interaction of a lump with a specified mean flow. 

\subsection{Two-dimensional reductions of the modulation equations}

In this subsection, we consider the (1+1)- and (2+0)-dimensional reductions of the full modulation system~\eqref{e:mod_full} that are obtained when one can neglect variations with respect to one of the independent variables.
Moreover, we show that each reduction can be diagonalized in terms of Riemann invariants. 

\paragraph{$XT$-dependent reduction.}
First, in the case of modulations that are independent of $Y$, we have that $\v$ is constant, and \eqref{e:mod_full} reduces to the three-component system
\bse
\begin{align}
\label{e:mod_X_a}
    a_{T}+(\u+a^2+q^2)a_{X}+a\u_{X}=0,\\
    \label{e:mod_X_aq}
    (aq)_{T}+(\u+a^2+q^2)(aq)_{X}=0,\\
    \label{e:mod_X_u}
    \u_{T}+\u\u_{X}=0.
\end{align}
\ese
or, in matrix form,
\begin{equation}
\label{e:mod_X}
\tilde{\@v}_{T}+\tilde{\mathsf{A}}\tilde{\@v}_{X}=0,\qquad\tilde{\@v}=\begin{bmatrix}
        a\\q\\\u
    \end{bmatrix},\qquad
    \tilde{\mathsf{A}}=\begin{bmatrix}
        \u+a^{2}+q^{2}&0&a\\
        0&\u+a^2+q^2&-q\\
        0&0&\u
    \end{bmatrix}.
\end{equation}
The eigenvalues of the matrix $\tilde{\mathsf{A}}$ are $\lambda_{1}=\lambda_{2}=\u+a^2+q^2$ and $\lambda_{3}=\u$ with respective 
left eigenvectors
\begin{equation}
\@l_{1}=\begin{bmatrix}  q\\a\\0 \end{bmatrix},\qquad
\@l_{2}=\begin{bmatrix} {(a^2+q^2)}/{a}\\0\\1 \end{bmatrix},\qquad
\@l_{3}=\begin{bmatrix} 0\\0\\1 \end{bmatrix}.
\end{equation}
From \eqref{e:mod_X_aq}, we have that $R_{1}=aq$ is a Riemann invariant with characteristic speed $\u+a^2+q^2$. 
(This is also apparent from $\d R_{1}=\@l_{1}\cdot \d\tilde{\@v}$.) 
The other Riemann invariant with the same characteristic speed is found from $\d R_{2}=\@l_{2}\cdot \d\tilde{\@v}$, which yields
\begin{equation}
    R_{2}=\int\frac{a^2+q^2}{a}\,\d a+\u.
\end{equation}
Since $q=R_{1}/a$, we find
\begin{equation}
    R_{2}=\frac{1}{2}a^2-\frac{R_{1}^{2}}{2a^{2}}+\u=\frac{a^{2}-q^2}{2}+\u.
\end{equation}
Finally, from \eqref{e:mod_X_u} the third Riemann invariant is $R_{3}=\u$. 
In terms of the Riemann invariants, the characteristic speeds are given by
\begin{equation}
    \lambda_{1}=\lambda_{2}=R_{3}+2\sqrt{(R_{2}-R_{3})^{2}+R_{1}^{2}},\qquad\lambda_{3}=R_{3}.
\end{equation}
Thus, the diagonal form of \eqref{e:mod_X} can be written compactly as
\begin{equation}
    \label{e:mod_X_diag}
    \partial_{T}R_{i}+\lambda_{i}\partial_{X}R_{i}=0,\qquad i=1,2,3.
\end{equation}
Note the system~\eqref{e:mod_X} is not genuinely nonlinear nor strictly hyperbolic, since two of the characteristic speeds coincide.

\paragraph{$YT$-dependent reduction.}
In the case of modulations that are independent of $X$, there is no reduction in the number of dependent variables, but the system \eqref{e:mod_full} reduces to
\vspace*{-0.4ex}
\bse
\label{e:mod_Y_full}
\begin{align}
\label{e:mod_Y_a}
    a_{T}-2qa_{Y}=0,\\
    \label{e:mod_Y_q}
    q_{T}-2qq_{Y}+\u_{Y}=0,\\
    \label{e:mod_Y_u}
    \u_{T}-\v_{Y}=0,\\
    \label{e:uy0}
    \u_{Y}=0.
\end{align}
\ese
Using \eqref{e:uy0} in \eqref{e:mod_Y_q}, we see that the lump parameters are fully decoupled from the mean field in this reduction. First considering only \eqref{e:mod_Y_u} and \eqref{e:uy0}, it can be deduced that 
\begin{equation}
    \u=f(T),\qquad\v=f'(T)Y+g(T),
\end{equation}
for some functions $f$ and $g$ of only $T$. Then, the remaining system shows that the lump parameters satisfy
\begin{equation}\label{e:YTsystem}
    \begin{bmatrix}
        a\\q
    \end{bmatrix}_{T}+\begin{bmatrix}
        -2q&0\\0&-2q
    \end{bmatrix}\begin{bmatrix}
        a\\q
    \end{bmatrix}_{Y}=0,
\end{equation}
which is already in diagonal form.
Similarly to the $XT$-reduction~\eqref{e:mod_X}, the system~\eqref{e:YTsystem} is not genuinely nonlinear nor strictly hyperbolic, since the two characteristic speeds coincide.

\paragraph{$XY$-dependent reduction.}
Finally, we consider the case of $T$-independent modulations, in which the system can be written as
\begin{equation}\label{e:XYsystem}
    \@v_{Y}+\mathsf{C}\@v_{X}=0,\qquad\mathsf{C}=\mathsf{B}^{-1}\mathsf{A}=\begin{bmatrix}
        -\frac{\u+a^2+q^2}{2q}&0&-\frac{a}{2q}&0\\
        0&-\frac{\u+a^2+q^2}{2q}&\frac{1}{2}&-\frac{1}{2q}\\
        0&0&0&-1\\
        0&0&-\u&0
    \end{bmatrix}.
\end{equation}
The eigenvalues of $\mathsf{C}$ are found to be
\begin{equation}
    \lambda_{1}=\lambda_{2}=-\frac{\u+a^2+q^2}{2q},\qquad\lambda_{3}=\sqrt{\u},\qquad\lambda_{4}=-\sqrt{\u},
\end{equation}
with corresponding left eigenvectors
\begin{equation}
\@l_{1}=\begin{bmatrix}
    a^2-q^2+\u\\
    -2aq\\
    a\\0
\end{bmatrix},\qquad\@l_{2}=\begin{bmatrix}
    2aq\\a^2-q^2+\u\\q\\1
\end{bmatrix},\qquad\@l_{3}=\begin{bmatrix}
    0\\0\\-\sqrt{\u}\\1
\end{bmatrix},\qquad\@l_{4}=\begin{bmatrix}
    0\\0\\\sqrt{\u}\\1
\end{bmatrix}.
\end{equation}
The left eigenvectors spanning the eigenspace of the double eigenvalue have been chosen such that it is straightforward to determine the related Riemann invariants by integration. In particular, from $\d R_{1}=\@l_{1}\cdot\d \@v$ and $\d R_{2}=\@l_{2}\cdot\d \@v$ we find
\begin{equation}
\label{e:XY_R12}
    R_{1}=\frac{1}{3}a^3-aq^2+a\u,\qquad R_{2}=a^{2}q-\frac{1}{3}q^{3}+q\u+\v.
\end{equation}
Note that $R_{1}$ is the Hamiltonian $\mathcal{H}$ as in \eqref{e:Hamiltonian}, which is generically a conserved quantity of the full modulation system when the mean field is time-independent. The Riemann invariants corresponding to $\pm\sqrt{\u}$ are found to be
\begin{equation}
\label{e:XY_R34}
    R_{3}=\frac{2}{3}\u^{3/2}-\v,\qquad R_{4}=-\frac{2}{3}\u^{3/2}-\v.
\end{equation}
From \eqref{e:XY_R34}, we can express the corresponding characteristic speeds as
\begin{equation}
    \lambda_{3}=\left[\frac{3}{4}(R_{3}-R_{4})\right]^{1/3},\qquad\lambda_{4}=-\left[\frac{3}{4}(R_{3}-R_{4})\right]^{1/3}.
\end{equation}
On the other hand, to determine the double characteristic speed $\lambda_{1}=\lambda_{2}$ in terms of the Riemann invariants, one needs to invert the cubic relations in \eqref{e:XY_R12}, which is possible but involves cumbersome formulae. The diagonal system can then be expressed as
\begin{equation}
    \partial_{Y}R_{i}+\lambda_{i}\partial_{X}R_{i}=0,\qquad i=1,2,3,4.
\end{equation}
Note that, once again, the system~\eqref{e:XYsystem} is not genuinely nonlinear or strictly hyperbolic, because two of the characteristic speeds coincide. 

We should briefly comment on how the system~\eqref{e:XYsystem} should be interpreted,  
since time-independent equations may at first appear to be incongruous with the lump being inherently a time-dependent entity. 
One should realize, however, that \eqref{e:XYsystem} describes the \emph{modulations} of a lump field
rather than simply an individual lump (see also the comment at the end of section~\ref{s:derivation}).
It is therefore entirely consistent for such modulations to be stationary even though a single lump propagating along the characteristic curves experiences changes over time. 
In this respect, the situation is perfectly analogous to what happens with the reductions of the soliton Whitham modulation equations studied in \cite{RHB2021,RMHB2021},
where it was demonstrated that, even though the modulation equations are independent of one of the coordinate variables, the corresponding dynamics for the soliton still involves changes over that coordinate. 

\section{Interactions between lumps and a rarefaction wave}
\label{s:lumpmeaninteractions}

The $Y$-independent reduction of the dKP equation is the familiar Hopf equation, $\u_t + \u \u_x = 0$, which admits the rarefaction wave solution 
\vspace*{-0.4ex}
\begin{equation}\label{e:HopfRW}
    \u(X,T)=\begin{cases}
        0,\quad&X<0\\
        X/T,\quad&0<X<T\\
        1,\quad&X>T
    \end{cases},
\end{equation}
corresponding to the initial condition $\u(X,Y,0)=\vartheta(X)$ for the dKP equation, where $\vartheta(X)$ is the Heaviside step function. 
Thus, \eqref{e:HopfRW} is also a $Y$-independent solution of the dKP equation \eqref{e:dKP}.
The bottom value can be taken to be zero without loss of generality due to the Galilean invariance of the KP equation, and the height of the step can be taken to be $1$ without loss of generality due to the scaling invariance of the KP equation. 
We should note that the case of an initial step placed along a line $X = c Y$ with $c\ne 0$ instead of $X=0$
[corresponding to an initial condition $u(X,Y,0) = \vartheta(X - c Y)$] would result in a slanted rarefaction wave.
On the other hand, this scenario can always be reduced to the above scenario using the invariance of the KP equation under a slanted change of the coordinate axes (e.g., see \cite{ABW2017}).
We can therefore limit ourselves to considering $c =0$ without loss of generality.

\begin{figure}[t!]
\kern-1ex
\centerline{\includegraphics[width=0.50\textwidth]{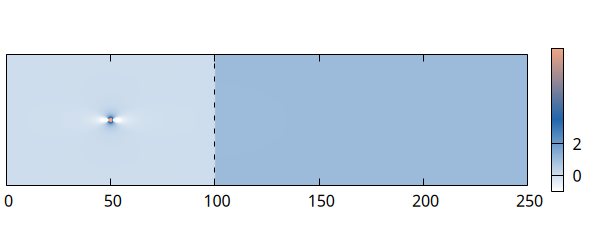}
    \includegraphics[width=0.50\textwidth]{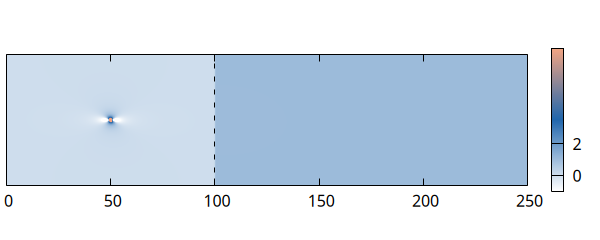}\kern-1em}
\kern-1ex
\centerline{\includegraphics[width=0.50\textwidth]{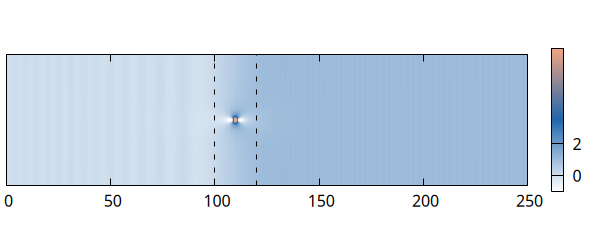}
    \includegraphics[width=0.50\textwidth]{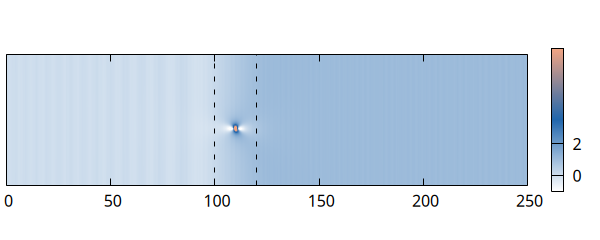}\kern-1em}
\kern-1ex
\centerline{\includegraphics[width=0.50\textwidth]{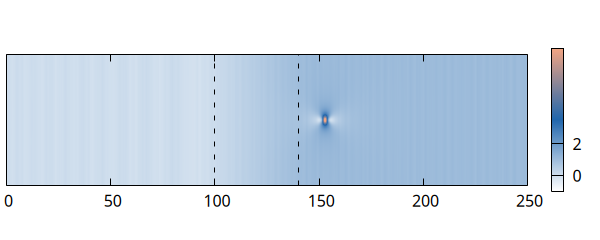}
    \includegraphics[width=0.50\textwidth]{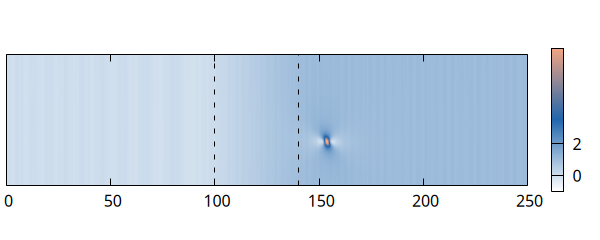}\kern-1em}
\kern-1ex
\centerline{\includegraphics[width=0.50\textwidth]{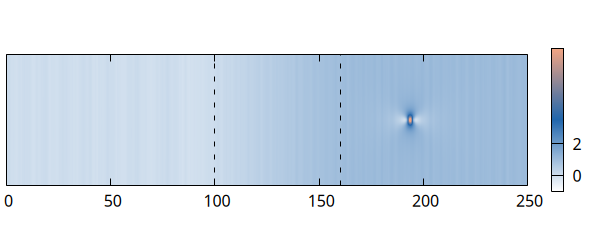}
    \includegraphics[width=0.50\textwidth]{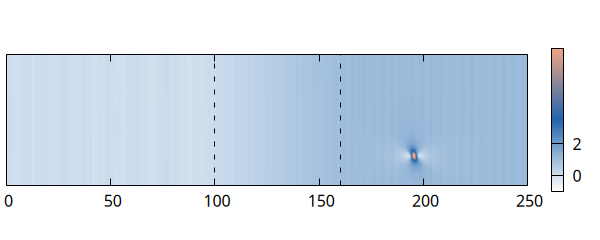}\kern-1em}
\kern-1ex
\centerline{\includegraphics[width=0.496\textwidth]{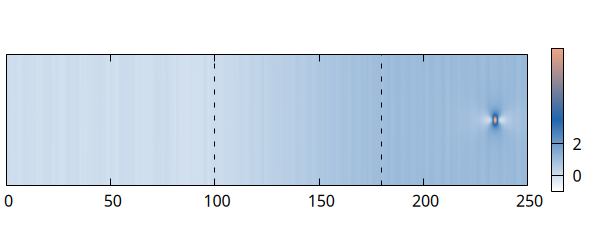}
    \includegraphics[width=0.496\textwidth]{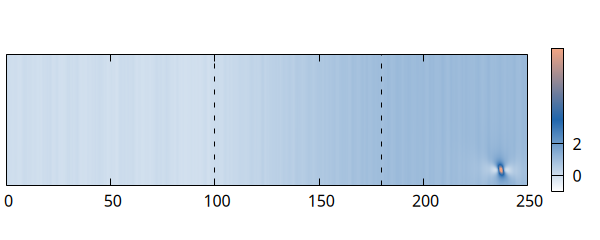}\kern-1em}
\smallskip
\caption{Density plots of the numerically computed solution $u(x,y,t)$ of the KPI equation as a function of $x$ and $y$ at the times $t = 0$, $t = 20$, $t = 40$, $t = 60$, and $t=80$ (top to bottom) with $a=1.75$ and $q=0$ (left) and $q=0.1$ (right). The rarefaction wave region lies between the two dashed vertical lines.
(See Appendix for details of the numerical methods.)}
\label{f:dens_1}
\end{figure}

If a lump is initially placed at a location $x_{0}<0$ to the left of the step, 
its horizontal velocity is positive by \eqref{e:lumpphases}.
Thus, 
the lump will eventually reach the "ramp" region of the rarefaction wave.
In this section we are interested in studying the dynamics of such a lump as it travels up the ramp.
We do so by making use of the $XT$-dependent modulation system \eqref{e:mod_X}. 
An illustration of this scenario is provided in Fig.~\ref{f:dens_1}, in which the left and right columns show temporal slices corresponding respectively to the cases of a horizontal and non-horizontal lump velocity.
(See the appendix for a description of the numerical methods used.)
Next, we consider these two cases separately, since they give rise to qualitatively different results.

\subsection{Interactions between a horizontally traveling lump and a rarefaction wave}

We begin with the simpler scenario $q=0$, in which case the velocity of the lump is oriented purely in the horizontal direction. 
In particular, \eqref{e:lumpphases} yields the initial horizontal velocity of the lump as $a_{0}^{2}$, where $a_{0}$ is the initial value of the amplitude parameter $a$. 
If the lump is transmitted through the rarefaction wave, it will emerge after some time from the ramp 
with an amplitude $a_1$ to be determined
and propagate (on top of the unit mean) with a constant velocity of $a_{1}^2+1$. 

To determine $a_{1}$ in terms of $a_{0}$, one can use the Riemann invariant of the $XT$-dependent modulation system $R_{2}=a^2/2+\u$. In particular, from \eqref{e:mod_X_diag} we know that this Riemann invariant is constant along the characteristics defined by $\d X/\d T=\lambda_{2}$ where $\lambda_{2}=\u+a^2$ is exactly the lump velocity. That is, $R_{2}$ is constant on the trajectory of the lump. 
As such, the values of $R_{2}$ before and after the interaction with the rarefaction ramp must be equal, which implies that
\vspace*{-0.6ex}
\begin{equation}
\label{e:a1}
    a_{1}=\sqrt{a_{0}^{2}-2}.
\end{equation}
Note that this relationship suggests that if $a_{0}<\sqrt{2}$, the lump will never emerge from the ramp.
We therefore have the following dichotomy:
\vspace*{-0.6ex}
\begin{enumerate}[label=(\roman*)]
\item
$a_{0}>\sqrt{2}$ results in lump transmission,
\item
$a_{0}<\sqrt{2}$ results in lump trapping.
\end{enumerate}
This result is similar to the transmission condition found for the KdV soliton-rarefaction wave interaction in \cite{ACEHL2023}. 
As we show next, further information about the dynamics of the lump within the ramp region, including the precise temporal dependence of the lump amplitude as it travels through the rarefaction wave, can be obtained by explicitly solving the modulation system \eqref{e:characteristics} or, equivalently in this case, the reduction~\eqref{e:mod_X_a}.

In the $XT$-dependent reduction with $q=0$, 
\eqref{e:mod_X_aq} is satisfied trivially.
Moreover, once a solution for~\eqref{e:mod_X_u} is given,
\eqref{e:mod_X_a} yields, along characteristics,
\begin{equation}
    \frac{\d a}{\d T}=-a\u_{X},\qquad\frac{\d X}{\d T}=\u+a^2.
\label{e:dadT_case1}
\end{equation}
Since the mean field at the bottom of the ramp is zero and the lump is initially centered at the location $x_0<0$, \eqref{e:lumpphases} yields
the time at which the lump reaches the base of the ramp as $T_{0}=-x_{0}/a_{0}^2$. 
In the ramp region, we can substitute the self-similar solution $\u=X/T$ from~\eqref{e:HopfRW}, in which case the solution of~\eqref{e:dadT_case1} subject to the initial conditions $a(T_{0})=a_{0}$, $X(T_{0})=0$ is given by
\begin{equation}\label{e:a(T)}
    a(T)=\frac{a_{0}T_{0}}{T},\qquad X(T)=\frac{a_{0}^{2}}{2}T\left(1-\frac{T_{0}^{2}}{T^{2}}\right).
\end{equation}
Now, if the lump is to transmit through the rarefaction wave, there must exist a time $T_{1}>T_{0}$ such that $X(T_{1})=T_{1}$ is satisfied, which yields a condition that can be used to solve for $T_1$.  
With the form of $X(T)$ given above, the only positive solution of this equation is 
\begin{equation}
\label{e:T1}
    T_{1}=\frac{a_{0}T_{0}}{\sqrt{a_{0}^{2}-2}}.
\end{equation}
Note that $T_{1}$ is not real-valued when $a_{0}<\sqrt{2}$, consistent with the previously found trapping condition. 
Conversely, if $T_{1}$ is real-valued, then the final value of the amplitude parameter after the lump emerges from the ramp is 
\begin{equation}
    a_{1}=a(T_{1})=\frac{a_{0}T_{0}}{T_{1}},
\end{equation}
which coincides with \eqref{e:a1} after substituting \eqref{e:T1}.

Recall that the physical amplitude $A$ of the lump is related to the parameter $a$ via $A=8a^2$. Summarizing the present results in terms of the amplitude, we then have that if the initial amplitude of the lump satisfies the condition $A_{0}>16$, then the lump is transmitted through the rarefaction wave with a final amplitude of $A_{1}=A_{0}-16$ on top of the unit mean field. Conversely, in the case where $A_{0}<16$, the lump will remain trapped in the ramp forever, never reaching the top of the rarefaction wave, and its amplitude will asymptotically decay to zero as $T\to\infty$, similarly to what happens with soliton trapping in the KdV equation \cite{ACEHL2023}.

Note that the full trajectory of a transmitting lump for all time is described by
\begin{equation}
    X_{\text{lump}}(T)=\begin{cases}
        x_{0}+a_{0}^{2}T,\quad&T\leq T_{0}\\
        \frac{a_{0}^{2}}{2}T\left(1-\frac{T_{0}^{2}}{T^{2}}\right),\quad& T_{0}<T<T_{1}\\
        x_{1}+(a_{1}^{2}+1)T,\quad&T\geq T_{1}
    \end{cases}.
\end{equation}
Using the explicit expressions \eqref{e:a(T)} shows that $x_{1}=x_{0}a_{1}/a_{0}$, i.e. the lump emerges from the interaction with a phase shift given by
\begin{equation}
\label{e:phase}
    \delta=x_{1}-x_{0}=\left(\frac{a_{1}}{a_{0}}-1\right)x_{0}.
\end{equation}
This phase shift can also be determined without knowledge of the full trajectory. To this end, we introduce an equally spaced array of horizontally traveling lumps with wavenumber $k$ and couple an approximate conservation of waves equation to the modulation system,
\begin{equation}
    \label{e:waves}
    k_{T}+\big[\big(2R_{2}-\u\big)k\big]_{X}=0,
\end{equation}
where $2R_{2}-\u=a^2+\u$ is the lump velocity. Since $R_{2}$ is constant, \eqref{e:waves} is diagonalized by the Riemann invariant $R_{k}$ given by
\begin{equation}
    \partial_{T}R_{k}+(2R_{2}-\u)\partial_{X}R_{k}=0,\qquad R_{k}=k\sqrt{R_{2}-\u}.
\end{equation}
The constancy of this Riemann invariant along the lump trajectory yields 
\begin{equation}
    k_{1}\sqrt{R_{2}-1}=k_{0}\sqrt{R_{2}},
\end{equation}
where $k_{0}$ and $k_{1}$ are the wavenumber (i.e. lump spacing) before and after the interaction, respectively. The phase shift can then be predicted by examining the ratio of the spacing before and after the interaction in the limit as $k\rightarrow0$,
\begin{equation}
    \frac{x_{1}}{x_{0}}\sim\frac{k_{0}}{k_{1}}\sim\sqrt{\frac{R_{2}-1}{R_{2}}}=\frac{a_{1}}{a_{0}},
\end{equation}
which is consistent with \eqref{e:phase}.

\begin{figure}[t!]
\smallskip
\centering
    \includegraphics[width=0.328\textwidth]{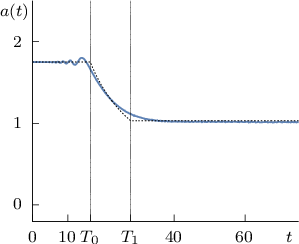}
    \includegraphics[width=0.328\textwidth]{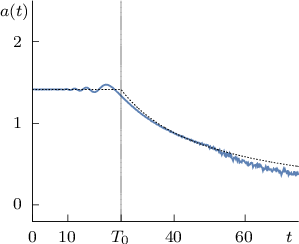}
    \includegraphics[width=0.328\textwidth]{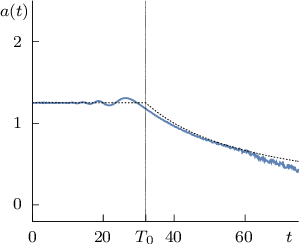}
    \caption{Numerically measured lump amplitude parameter, $a(t)$ as a function of $t$ (solid blue line) with theoretical prediction (dashed black line) in a case of lump transmission, $a_0=1.75$, $q_0 = 0$ (left), critical trapping threshold $a_0 = \sqrt{2}$, $q=0$ (center), and below the critical threshold $a_0 = 1.25$, $q=0$ (right).}
\label{f:amplitude_1par}
\end{figure}

Figure \ref{f:amplitude_1par} displays a comparison between the analytical prediction \eqref{e:a(T)} for the amplitude parameter and the numerically measured amplitude of the lump throughout the simulation.
In particular, the left panel corresponds to a case of lump transmission (i.e., $a_0>\sqrt2$), while the  right panel illustrates  the phenomenon of lump trapping (i.e., $a_0<\sqrt2$), in which the lump never emerges from the ramp and its amplitude decays for all time, and the center panel corresponds to the critical case (i.e., $a_0 = \sqrt2$). 
Here and in subsequent comparisons with numerics, we set $\e=1$ so that $T=t$, $X=x$, $Y=y$ and set $-x_0 \gg 1$.
Note how the analytical prediction for the final lump amplitude is in excellent agreement with the results from the numerical simulations.  (See Fig.~\ref{f:amp_oblique} for further confirmation of this agreement.) 
The only small disagreements between the theory and the numerics are the small oscillations of the lump amplitude before it enters the rarefaction wave and the smooth transition to the asymptotic value of the lump amplitude. 
Both of these phenomena are due to dispersive effects that are not captured by the ramp solution~\eqref{e:HopfRW}, which results from the reduction from the dispersive KP equation to the dispersionless Hopf equation.
On the other hand, the general features of the temporal dependence of the lump amplitude, and its asymptotic value are very well captured by the solution of the modulation equations.
(The smaller noise-like oscillations of the lump amplitude toward the end of simulation are likely due to corruption of the lump due to interference with the rest of the numerical solution field. 
See the Appendix for a discussion of the numerical methods.)

\subsection{Interactions between an obliquely traveling lump and a rarefaction wave}

We now move on to study the interaction between the rarefaction wave and a full two-parameter lump with nonzero transverse velocity. In this case, from \eqref{e:lumpphases} we have that, if the lump is initially placed at a location $(x_{0},y_{0})$ with $x_0<0$ and $y_{0}$ arbitrary, the time at which the lump reaches the base of the rarefaction ramp is $T_{0}=-x_{0}/(a_{0}^2+q_{0}^2)$. Interestingly, we find that in this case, the lump will always transmit through the ramp as long as $q_{0}\neq0$. We first show that this is the case by considering the Riemann invariants of the $XT$-dependent modulation system \eqref{e:mod_X} given by $R_{1}=aq$ and $R_{2}=(a^2-q^2)/2+\u$. Using similar logic as in the previous subsection, since $R_{1}$ and $R_{2}$ are both constant on the trajectory of the lump, it must be the case that
\vspace*{-0.6ex}
\bse
\begin{align}
\label{e:aq1}
a_{1}q_{1}&=a_{0}q_{0},\\
\label{e:aq2}
    a_{1}^2-q_{1}^2&=a_{0}^2-q_{0}^{2}-2.
\end{align}
\ese
 Using \eqref{e:aq1} we can write $q_{1}=a_{0}q_{0}/a_{1}$ in \eqref{e:aq2} to get the biquadratic equation
\begin{equation}
    a_{1}^4-\mu a_{1}^2-a_{0}^2q_{0}^2=0,\qquad\mu=a_{0}^2-q_{0}^2-2,
\end{equation}
whose only positive solution is
\begin{equation}
\label{e:a1_2par}
    a_{1}^2=\frac{1}{2}\left(\mu+\sqrt{\mu^2+4a_{0}^2q_{0}^2}\right).
\end{equation}
Note that in the one-parameter case $q_{0}=0$, the sign of $\mu$ distinguishes between the cases of trapping and transmission. However, in the two-parameter case $q_{0}\neq0$, the final amplitude parameter $a_{1}$ is well-\unskip defined regardless of the sign of $\mu$. 

\begin{figure}[t!]
\centering
    \includegraphics[width=0.328\textwidth]{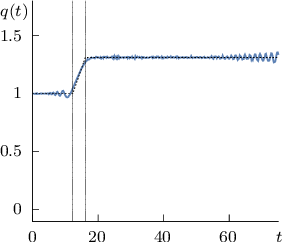}
    \includegraphics[width=0.328\textwidth]{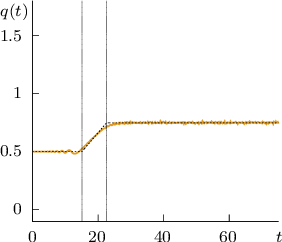}
    \includegraphics[width=0.328\textwidth]{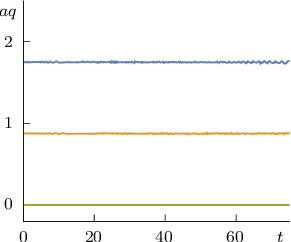}
    \vskip0.4\smallskipamount
    \caption{The lump parameter $q(t)$ as a function of time as the lump propagates  through the rarefaction wave, with $a_0=1.75$ and with $q_0=1$ (left panel) and $q_0=0.5$ (center panel) as reconstructed from numerical simulations (blue and gold curves),
    compared to the theoretical predictions (dashed black curves).
    Right panel: The product $aq$ as function of time in three simulations with $a_0=1.75$ and $q_0=0$ (green curve), $q_0=0.5$ (gold curve) and $q_0=1$ (blue curve). }
\label{f:aq_const}
%\end{figure}
\vskip2\bigskipamount
%\begin{figure}[h!]
\centering
    \includegraphics[width=0.328\textwidth]{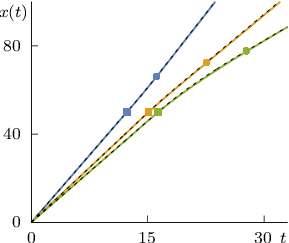}
    \includegraphics[width=0.328\textwidth]{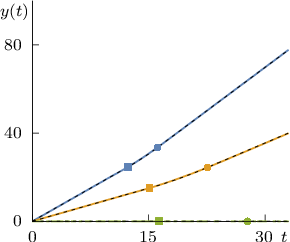}
    \includegraphics[width=0.328\textwidth]{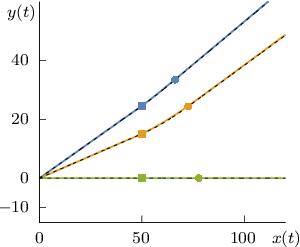}
    \vskip0.4\smallskipamount
    \caption{The $x$-coordinate of the lump (left panel), the $y$-coordinate of the lump (center panel), and the lump trajectory $(x(t),y(t))$ (right panel) for $a=1.75$ and $q=0$ (green), $q=0.5$ (gold) and $q=1$ (blue) vs with theory (dashed black line). 
    The squares mark the time at which the lump enters the rarefaction wave region, 
    while the circles mark the time at which the lump exits the rarefaction wave.}
\label{f:trajectory}
\end{figure}

The full trajectory of the lump as it interacts with the rarefaction wave can be found by solving \eqref{e:characteristics}. The evolution of the amplitude parameter $a$ is the same as in the previous subsection, since the first equation in \eqref{e:characteristics} applies in either case. 
Moreover, since $aq$ is a constant of motion for a $Y$-independent mean field, we have that $q$ grows with time as $a$ decays. That is, explicitly,
\begin{equation}\label{e:q(T)}
    a(T)=\frac{a_{0}T_{0}}{T},\qquad q(T)=\frac{q_{0}T}{T_{0}}.
\end{equation}
The time-independence of the product $aq$ throughout the interaction is confirmed by the numerical results, as shown in Fig.~\ref{f:aq_const}. 
The temporal dependence of $q(T)$ is also shown in Fig.~\ref{f:aq_const} in two representative cases and compared to the theoretical predictions from \eqref{e:q(T)}, again showing very good agreement.
(As before, we suspect that the small noise-like oscillations of $q$ toward the end of simulation are due to corruption of the lump due to interference with the rest of the numerical solution field.)
Note that the growth of $q$ is the reason that a lump with nonzero transverse momentum is always transmitted through a rarefaction wave: Since the horizontal component of the lump velocity is $\u+a^2+q^2$, the fact that $q$ grows with time shows that the lump experiences \textit{acceleration} in the horizontal direction as a result of the interaction. 

Finally, we can solve the modulation equations for the characteristic trajectories to find
\vspace*{-0.5ex}
\begin{equation}
\label{e:traj}
    X(T)=\frac{a_{0}^{2}}{2}T\left(1-\frac{T_{0}^{2}}{T^{2}}\right)-\frac{q_{0}^{2}}{2}T\left(1-\frac{T^{2}}{T_{0}^{2}}\right),\qquad  Y(T)=q_{0}T_{0}\left(1-\frac{T^{2}}{T_{0}^{2}}\right)+Y(T_{0}),
\end{equation}
The time at which the lump exits the ramp is $T_{1}$ such that $X(T_{1})=T_{1}$. This is equivalent to
\begin{equation}
    q_{0}^{2}T_{1}^{4}+\mu T_{0}^{2}T_{1}^2-a_{0}^{2}T_{0}^{4}=0.
\end{equation}
The relevant solution is
\begin{equation}
    T_{1}=\frac{T_{0}}{\sqrt{2}|q_{0}|}\sqrt{-\mu+\sqrt{\mu^2+4a_{0}^{2}q_{0}^2}},
\end{equation}
which is defined for either sign of $\mu$ as long as $q_{0}\neq0$. 
Note that taking the limit of the above expression as $q_{0}\rightarrow0$ recovers \eqref{e:T1}. With this expression for $T_{1}$, $a_{1}=a(T_{1})$ is identical to the previously obtained prediction \eqref{e:a1_2par}. 
Figure~\ref{f:trajectory} shows the theoretical predictions in \eqref{e:traj} for the lump coordinates in a number of representative cases, compared with the results of corresponding numerical simulations.
The agreement between theory and numerical results is excellent in all cases considered.

Summarizing, in terms of the physical lump amplitude $A=8a^{2}$ we have that any lump with initial amplitude $A_{0}$ and $q_{0}\neq0$ is expected to emerge from the rarefaction wave with new amplitude given by
\vspace*{-0.6ex}
\begin{equation}\label{e:finalamplitude2}
    A_{1}=4\left(\mu+\sqrt{\mu^2+\frac{1}{2}A_{0}q_{0}^2}\right),\qquad\mu=\frac{1}{8}A_{0}-q_{0}^2-2.
\end{equation}
The trajectory of the lump for all time is
\bse
\begin{align}
    X_{\text{lump}}(T)&=\begin{cases}
        x_{0}+(a_{0}^{2}+q_{0}^{2})T,\quad&T\leq T_{0}\\
        \frac{a_{0}^{2}}{2}T\left(1-\frac{T_{0}^{2}}{T^{2}}\right)-\frac{q_{0}^{2}}{2}T\left(1-\frac{T^{2}}{T_{0}^{2}}\right),\quad& T_{0}<T<T_{1}\\
        x_{1}+(a_{1}^{2}+q_{1}^{2}+1)T,\quad&T\geq T_{1}
    \end{cases},\\
    Y_{\text{lump}}(T)&=\begin{cases}
        y_{0}-2q_{0}T,\quad&T\leq T_{0}\\
        q_{0}T_{0}\left(1-\frac{T^{2}}{T_{0}^{2}}\right)+y_{0}-2q_{0}T_{0},\quad& T_{0}<T<T_{1}\\
        y_{1}-2q_{1}T,\quad&T\geq T_{1}
    \end{cases}.
\end{align}
\ese
By matching at $T=T_{1}$ we find 
\begin{equation}
    x_{1}=\frac{x_{0}T_{1}}{T_{0}}\frac{a_{1}^2+q_{1}^2}{a_{0}^{2}+q_{0}^{2}},\qquad y_{1}=y_{0}-q_{0}T_{0}\left(1-\frac{T_{1}^{2}}{T_{0}^{2}}\right).
\end{equation}
This first equation above shows that the phase shift in $x$ as a result of the interaction is 
\begin{equation}
\label{e:phase2}
    \delta=x_{1}-x_{0}=\left(\frac{T_{1}}{T_{0}}\frac{a_{1}^{2}+q_{1}^{2}}{a_{0}^{2}+q_{0}^{2}}-1\right)x_{0},
\end{equation}
which reduces to \eqref{e:phase} when $q_{0}=q_{1}=0$. Similarly to the previous case, this phase shift can also be obtained by considering the conservation of waves type equation
\begin{equation}
    k_{T}+\left[\left(\u+2\sqrt{(R_{2}-\u)^{2}+R_{1}^{2}}\right)k\right]_{X}=0,
\end{equation}
with $R_{1}$ and $R_{2}$ held constant. This is diagonalized by the Riemann invariant
\begin{equation}
    R_{k}=kp(\u),\qquad p(\u)=\frac{(R_{2}-\u)\left(R_{2}-\u+\sqrt{(R_{2}-\u)^2+R_{1}^2}\right)+R_{1}^{2}}{\left(R_{2}-\u+\sqrt{(R_{2}-\u)^2+R_{1}^2}\right)^{3/2}}.
\end{equation}
Using the fact that $R_{k}$ remains constant throughout the interaction yields
\begin{equation}
    \frac{x_{1}}{x_{0}}\sim\frac{k_{0}}{k_{1}}\sim\frac{p(1)}{p(0)}.
\end{equation}
Direct calculations show that this provides the same phase shift as in \eqref{e:phase2}.

\begin{figure}[t!]
%\smallskip
\centering
    \includegraphics[width=0.475\textwidth]{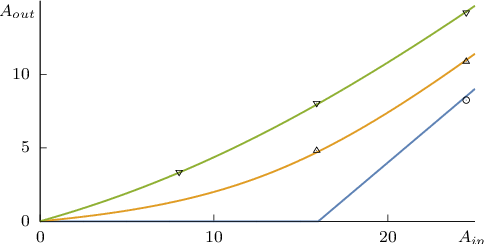} \quad
    \includegraphics[width=0.475\textwidth]{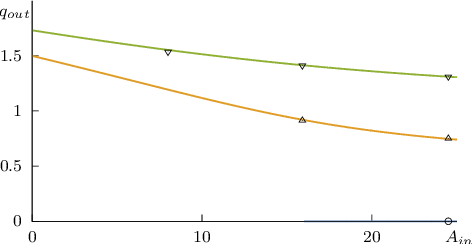}
    \smallskip
    \caption{Left: Analytical prediction for the final lump amplitude $A_\mathrm{out} = 8a_1^2$ as a function of the initial lump amplitude $A_\mathrm{in} = 8a_0^2$ for $q_0 = 0$ (blue curve), $q_0=0.5$ (gold curve) and $q_0=1$ (green curve) versus the numerically recovered values, marked by circles and triangles. 
    Right panel: analytical prediction for the final slope parameter $q_1$ as a function of the initial lump amplitude $A_\mathrm{in} = 8a_0^2$ (\unskip for the same values of~$q_0$ as in the left panel) versus the numerically recovered values.}
    \label{f:amp_oblique}
\end{figure}

In Fig.~\ref{f:amp_oblique}, the predicted values of the parameters of the \textit{transmitted} lump are plotted as a function of the amplitude of the \textit{initial} lump for several values of $q_{0}$.
The theoretical predictions are then compared with the numerically measured values from the corresponding numerical simulations.
In all cases, the agreement between theoretical predictions and numerical results is excellent. 

\section{Discussion and outlook}
\label{s:conclusions}

In this work we derived a four-component (2+1)-dimensional hyperbolic system of equations governing the modulations of a lump solution of the KPI equation in the presence of a slowly varying mean field. 
We showed that the system passes the Haantjes tensor test for integrability, admits a Hamiltonian formulation when the mean field is stationary, and we diagonalized it in terms of Riemann invariants in each of its three two-dimensional reductions. 
As a concrete application, we used the modulation system to derive sharp transmission and trapping criteria for a lump traversing a mean-field rarefaction wave: a horizontally-traveling lump is transmitted if and only if its initial amplitude exceeds the critical value $A_0 = 16$, while a lump with any nonzero transverse parameter $q_0$ is always transmitted, with the transverse momentum $aq$ conserved throughout the interaction. 
The analytic predictions were compared with direct numerical simulations of the KPI equation and 
found to be in excellent agreement. 

Although this work was motivated by recent extensions of Whitham modulation theory to the line-soliton and one-phase periodic solutions of the KP equation~\cite{ABW2017}, it differs from \cite{ABW2017} in one essential respect. 
Line solitons and one-phase periodic solutions are co-dimension-one objects (i.e., the solution is constant along the wave front), and the corresponding modulation theory is built by averaging over a one-dimensional family of phases.
Lumps, by contrast, are localized in both spatial directions, and decay only algebraically at infinity, so the conventional phase average is unavailable. 
The derivation in Section~\ref{s:derivation} instead proceeds from integral identities for the lump itself, taken in the moving reference frame of the lump and at leading order in the modulation parameter~$\varepsilon$. 
To the best of our knowledge, the system~\eqref{e:mod_full} is therefore the first Whitham-type modulation theory for a genuinely two-dimensional, fully localized coherent structure of an integrable~PDE.

The trapping/transmission dichotomy obtained in Section~\ref{s:lumpmeaninteractions} places this
work within the broader research on soliton--mean field interactions that has emerged over the past decade 
in the context of dispersive hydrodynamics~\cite{ElHoefer2016}. 
For the Korteweg--de Vries equation, the analogous problem of a soliton traversing a rarefaction was analyzed in~\cite{ACEHL2023}, and the trapping condition $A_0 < 2$ was obtained there; closely related analyses of soliton--dispersive shock and soliton--periodic mean interactions have since been carried out for KdV and other
weakly dispersive hydrodynamic models in~\cite{BiondiniLottes,ElHoefer2016,MaidenEtAl}.
Indeed, the extent to which the modulation theory of a genuinely two-dimensional integrable coherent structure parallels the one-dimensional KdV theory~\cite{ACEHL2023} is perhaps striking: a Hamiltonian system on the parameter space, Riemann invariants in each coordinate reduction, and a sharp threshold separating transmission from trapping. 
At the same time, our results extend this program in an essential way: here the coherent structure is genuinely two-dimensional, so that nontrivial transverse dynamics (notably the conservation of $aq$ that drives the unconditional transmission of obliquely-traveling lumps) becomes available for the first time. 

We also emphasize that the quantitative predictions for the final lump amplitude are in principle testable in the same optical setting in which lumps were recently observed~\cite{DPBTC2026}, and we hope the present work will stimulate such experiments.

The fact that the modulation system~\eqref{e:mod_full} passes the Haantjes test, admits a Hamiltonian
form for stationary means, and reduces to diagonal (1{+}1)-systems with explicit Riemann invariants in all three coordinate-plane reductions strongly suggests that the system should be amenable to the full machinery of Tsarev's generalized hodograph method~\cite{Tsarev1991} 
and to the broader theory of integrable systems of hydrodynamic type~\cite{DubrovinNovikov,Ferapontov,Pavlov}.
Two questions arise naturally. First, can the full $(2{+}1)$-system be integrated by hydrodynamic-type techniques --- for example, by exhibiting four functionally independent commuting flows, or by embedding its reductions in a hydrodynamic chain in the sense of Pavlov? 
Second, all three (1{+}1)-dimensional reductions analyzed in section~\ref{s:integrability} fail to be strictly hyperbolic, with two characteristic speeds always coinciding. 
The geometric meaning of this degeneracy, and its consequences for shock formation in the modulation variables, deserve a more thorough analysis. 
(We note that analogous degeneracies appear in the line-soliton modulations of the
KPII equation~\cite{ABW2017,RHB2021}.)

A complementary perspective is provided by Krichever's algebro-geometric approach to averaging on integrable systems~\cite{KricheverAveraging}. 
Since lumps arise as further rational degenerations of singular-curve theta-function solutions of the KP equation \cite{Agostini,Little},
the modulation system~\eqref{e:mod_full} may admit an interpretation as a singular limit of the corresponding Krichever--Whitham equations for the periodic case. Establishing this connection rigorously, and using it to inform the modulations of higher-order rational solutions, is an attractive open problem.
At the same time, we should point out that the same problem is also still open for the one-phase KP-Whitham system derived in \cite{ABW2017}. 

Finally, a long-standing obstacle in the analysis of the KP equation is the absence of an effective inverse scattering theory for genuinely non-decaying initial data: the IST has been developed for decaying data~\cite{AF1983_1,AF1983_2,AS1981} and for perturbations of an exact lump or line soliton~\cite{BPP2006,BPPP2002,VA2002,W2025},
but not for backgrounds of the type considered here. 
We believe that the development of a viable IST for non-decaying initial conditions and the formulation of a suitable semiclassical limit that can derive the system~\eqref{e:mod_full} in a rigorous way is one of the most challenging and at the same time intriguing open problems suggested by this work.

We end the discussion by mentioning that we view this work as the initiation of a new research program dealing with modulation theory for two-dimensional localized structures. In addition to the open problems discussed above, there are many directions for further investigation; including modulations of multi-lump configurations and lump chains, lump interactions with more complex mean fields such as dispersive shock waves, and modulation theory for lump-like solutions of other physically relevant models.
We hope that the results of this work and the present discussion will stimulate further work on these topics.

\paragraph{Acknowledgment.}  
We thank Gennady El for many interesting discussions on related topics. 
G.B.\ was partially supported by the Simons Foundation under grant number SFI-MPS-TSM-00013369. The work of M.H. was supported by NSF Grant No. DMS-2306319.

\paragraph{Conflict of interest.}
The authors declare that they have no conflict of interest.

\addcontentsline{toc}{section}{Appendix}
\section*{Appendix}

In this appendix we provide a concise description of the numerical methods used to perform the numerical simulations discussed in the main text.

\paragraph{Numerical integration of the KP equation.}
The KP equation~\eqref{e:KP} is solved numerically by a pseudo-spectral method combined with a fourth-order exponential time-differencing in time~\cite{KT2005} by recasting the Fourier-transformed equation into the form
\begin{align}
    \partial_t \hat u_k = L_k \hat u_k + \hat N(u) 
    \label{e:ft_pde}
\end{align}
where $\hat u_k(t)$ is the Fourier coefficient of $u(x,y,t)$ with $k = (k_x,k_y)$,
with 
\be
L_k = 
    \begin{cases} \displaystyle
      ik_x^3 + i{k_y^2}/{k_x}\,\,&k_x\neq 0 \\
      0\,\, &\mbox{otherwise}
    \end{cases} 
\ee
and where the nonlinear term is
\begin{align}
    \hat N(u) = -ik_x \mathcal{F} \left[\frac{u^2}{2}\right],
\end{align}
$\mathcal{F}$ denoting the Fourier transform. 
Equation~\eqref{e:ft_pde} is 
recast as an integral equation,
\begin{align}
    \hat u_k(t+h) = e^{hL_k}\hat u_k + \int_{0}^{h} e^{L(h-s)}\hat N(u(t+s))ds
\end{align}
and the nonlinear term is approximated by a fourth order Runge-Kutta (RK4) method. The resulting method is often referred to as the ETDRK4~\cite{KT2005}. 
The accuracy of the time integration is verified by monitoring the four conserved integrals
\bse
\begin{align}
    M &= \iint_{\mathbb{R}^{2}} u dxdy, \qquad
    P = \iint_{\mathbb{R}^{2}} u^2 dxdy, \qquad
    Q = \iint_{\mathbb{R}^{2}} u v dxdy, \\
    E &= \iint_{\mathbb{R}^{2}} \left[u_x^2 + \sigma v_y^2 + \frac{1}{3}u^3\right] dxdy
\end{align}
\ese
where $v_x = u_y$. 
Over the course of all simulations reported, the largest numerical error was observed in $E$, with $\max(\Delta E/E) \sim 10^{-4}$.

\paragraph{Initial and boundary conditions, simulation parameters.}
The theoretical problem is posed on the infinite plane $(x,y)\in\mathbb{R}^2$.  
For the purposes of numerical simulation, we restrict the spatial variables to a computational box $(x,y)\in[0,L_x]\times[0,L_y]$ with periodic boundary conditions in both directions. 
Note that the Heaviside initial condition considered in section~\ref{s:lumpmeaninteractions} is not periodic with respect to~$X$. 
Moreover, the discontinuous transition between the two step values would result in the generation of 
spurious high-wavenumber components that would pollute the solution field.
To eliminate both of these problems,
we choose numerical initial conditions consisting of a symmetrized, regularized step function,
\vspace*{-0.1ex}
\begin{align}
    u_s(x;x_c,w,\delta) = \tfrac{1}{2}\left(1 - \tanh \left(f(x)/\delta\right) \right),
    \qquad
    f(x) = (x-x_s)^2 - w^2,
    \label{e:step}
\end{align}
where $x_s$ and $w$ controls the step center and its half-width, and $\delta$ is gives the characteristic width of the step transition smoothness, a discontinuous step corresponding to the limit $\delta\to0$. 
The evenness of $f(\cdot)$ with respect to $x_c$ ensures that the limits of \eqref{e:step} as $x\to\pm\infty$ coincide. 
A lump soliton $U(\cdot)$ with parameters $a$ and $q$ given by equation~\eqref{e:lump} is then added to the above step-like initial condition, so that the full initial condition is
\begin{align}
    u(x,y,t=0) = u_s(x; x_c, w, \delta) + U(x-x_l,y-y_l; a_o,q_o),
\end{align}
where the initial lump parameters are $x_l = 50$, $y_l = 93.75$ and the step parameters are $x_c=275$, $w = 175$ and $\delta=100$ on a computational domain up to $L_x = 500$ by $L_y = 125$.
(Note that the slow, algebraic decay of the lump solution leads to the need to use a large spatial domain.)
The intial lump parameters $a_o$ and $q_o$ differ among the various simulations (see Fig~\ref{f:amplitude_1par}--\ref{f:amp_oblique}). 
Up to $32768\times8192$ Fourier modes were used to resolve the solution on the numerical grid,
even though in some cases a much smaller number of Fourier modes was sufficient to accurately capture the solution. 
The computational were performed on the computers of the Center for Computational Research at the University at Buffalo.

\paragraph{Lump detection algorithm.}
In the course of the simulations, the lump propagates into the rarefaction wave, and needs to be detected and its parameters $a,q$ recovered. A postprocessing script is run on the numerical data for $u(x,y,t_n)$ for a set of time slices $t_n = n\tau$ where $\tau = T_{max}/M$ with $M=500$ and $T_{max} = 100$. At each time slice, a rough estimate for the $\max u(x,y,t_n)$ is determined by scanning the maximum on the numerical grid, 
\begin{align}
    (x_{i_\mathrm{max}}, y_{j_\mathrm{max}}) = \arg \max \limits_{(x_i,y_j)} u(x_i,y_j,t_n).
\end{align} 
Next, a local grid patch at $(x_i,y_j)$ with $i = i_\mathrm{max}-l,\ldots,i_\mathrm{max}+l$ and $j = j_\mathrm{max}-l,\ldots,j_\mathrm{max}+l$ is considered, with the parameter $l$ controlling the size of this local patch. 
Once the data-points $(x_i,y_j,u_{ij})$ are determined, a nonlinear least-squares fit to the lump soliton~\eqref{e:lump} is performed to determine five parameters,
$(x_\mathrm{max},y_\mathrm{max})$, $a$,$q$ and $\bar u$. 
The resulting lump parameters are then compared to the theoretical predictions in Figs~\ref{f:amplitude_1par}--\ref{f:amp_oblique}.

\end{document}